\documentclass[structabstract]{aa}  
\usepackage{graphicx}
\usepackage{lscape}
\usepackage{txfonts}
\usepackage{natbib,twoopt}
\usepackage[breaklinks=true]{hyperref} 
\bibpunct{(}{)}{;}{a}{}{,} 
\newcommandtwoopt{\citeads}[3][][]{\href{http://adsabs.harvard.edu/abs/#3}%
{\citealp[#1][#2]{#3}}} 
\newcommandtwoopt{\citepads}[3][][]{\href{http://adsabs.harvard.edu/abs/#3}%
{\citep[#1][#2]{#3}}} 
\newcommandtwoopt{\citetads}[3][][]{\href{http://adsabs.harvard.edu/abs/#3}%
{\citet[#1][#2]{#3}}} 
\newcommandtwoopt{\citeyearads}[3][][]%
{\href{http://adsabs.harvard.edu/abs/#3}{\citeyear[#1][#2]{#3}}}
\begin{document}
   \title{Radial velocities and metallicities from infrared \ion{Ca}{II} triplet spectroscopy of open clusters }

   \subtitle{Berkeley~26, Berkeley~70, NGC~1798, and NGC~2266\fnmsep\thanks{Based on observations made with the 2.5 Isaac Newton Telescope operated on the island of La Palma by the Isaac Newton Group in the Spanish Observatorio del Roque de los Muchachos of the Instituto de Astrof\'{\i}sica de Canarias.}}
   \titlerunning{Radial velocities and metallicities of open clusters} 

   \author{R. Carrera\inst{1,2}}

   \institute{Instituto de Astrof\'{\i}sica de Canarias, La Laguna, Tenerife, Spain\\
             \email{rcarrera@iac.es}
             \and
             Departamento de Astrof\'{\i}sica, Universidad de La Laguna, Tenerife, Spain\\
             }

   \date{Received September 15, 2110; accepted March 16, 2220}

 
  \abstract
   {Open clusters are ideal test particles for studying the formation and evolution of the Galactic disk. However, the number of clusters with information about their radial velocities and chemical compositions remains largely insufficient.}
   {We attempt to increase the number of open clusters with determinations of radial velocities and metallicities from spectroscopy.}
   {We acquired medium-resolution spectra (R$\sim$8000) in the region of the infrared \ion{Ca}{II}
triplet lines ($\sim$8500\AA) for several stars in four open clusters with the long-slit spectrograph IDS at
the 2.5m Isaac Newton Telescope, Roque de los Muchachos Observatory, Spain.
Radial velocities were obtained by cross-correlating the observed spectra with
those of two template stars. We used the relationships available in the literature
between the strength of infrared \ion{Ca}{II} lines and metallicities to derive
the metal content of each cluster.}
   {We provide the first spectroscopic determinations of radial velocities and
metallicities for the open clusters Berkeley~26, Berkeley~70, NGC~1798, and
NGC~2266. We obtain $\langle V_r \rangle$=68$\pm$12, -15$\pm$7, 2$\pm$10, and
-16$\pm$15 km s$^{-1}$ for Berkeley~26, Berkeley~70, NGC~1798, and NGC~2266,
respectively. For Berkeley~26 we derive a metallicity of [Fe/H]=-0.35$\pm$0.17
dex. Berkeley~70 has a solar metallicity of [Fe/H]=-0.01$\pm$0.14 dex,
while NGC~1798 has a slightly lower metal content of [Fe/H]=-0.12$\pm$0.07 dex. Finally, we derive a metallicity of [Fe/H]=-0.38$\pm$0.06 dex for NGC~2266.}
   {}


   \maketitle
%

\section{Introduction}

Open clusters (OCs) have been widely used to investigate the existence of trends
in the Galactic disk such as radial and vertical metallicity
gradients or an age-metallicity relationship
\citepads[e.g.][]{1979ApJS...39..135J,1980A&A....81..375P,1997AJ....114.2556T,
2002AJ....124.2693F,2003AJ....125.1397C,2010A&A...511A..56P,2011A&A...535A..30C}
.
However, our knowledge of OC properties
is still far from complete. Age and distance estimations, mainly obtained
from isochrone fitting, are available for about $\sim$70\% of the more than
2100 OCs known in our Galaxy \citepads{2002A&A...389..871D}\footnote{The updated
version of this catalog can be found at {\tt
http://www.astro.iag.usp.br/~wilton/}.}. However, radial velocities have been
determined only for a 24\% of them. The picture is even worse in the case of
chemical compositions, which have been obtained for only 9\%, mainly
by means of different studies in the Washington
\citepads[e.g.][]{2005MNRAS.363.1247P}, DDO
\citepads[e.g.][]{1977AJ.....82...35J,1999A&AS..134..301C}, Str\"omgren
\citepads[e.g.][]{2003A&A...403..937P}, UBV
\citepads[e.g.][]{1998A&AS..128..131C}, IR
\citepads[e.g.][]{1999A&A...343..825C,2000A&A...353..147V}, and Vilnius
\citepads[e.g.][]{2011BaltA..20...27B,2011BaltA..20....1Z} photometric systems,
often giving rise to
considerable differences among them and to those obtained from spectroscopy.

Reliable information about the chemical composition can be retrieved only
from spectroscopy. However, the acquisition of high-resolution spectra
(R$\geq$20000), which is the only
way to derive detailed abundances, needs large amounts of telescope time. This
together with the complexity of the associated analysis explain why this kind of
study has been performed for only 4\% of known OCs
\citepads{2011A&A...535A..30C}. The alternative is to perform low- and
medium-resolution spectroscopy, although these data
provide information only information about the metallicities and radial
velocities of
stars
\citepads[e.g.][]{1993A&A...267...75F,2002AJ....124.2693F,2007AJ....134.1298C,
2009MNRAS.393..272W}.

\begin{table*}[htb]
\caption{Observing logs and program star information.}
\label{obsstars}
\centering          
\renewcommand{\footnoterule}{}
\begin{scriptsize}
\begin{tabular}{l c c c c c c c c c c c c c c} 
\hline\hline
	Cluster & Star\footnote{Identification taken from WEBDA database.} & $\alpha_{2000}$ & $\delta_{2000}$ & V & V-I & B-V & V$_{r}$ & EW$_{8498}$ & EW$_{8542}$ & EW$_{8662}$ & $\Sigma$ Ca & t$_{exp}$ & S/N$^{tot}$ & Note\\
	 &  & (hrs) & (deg) & (mag) & (mag) &(mag) & (km s$^{-1}$) & (\AA) & (\AA) & (\AA) & (\AA) & (sec) &  &\\
            \hline
Be~26    & 1038 & 06:50:02.9 & +05:44:59 & 18.845 & 1.283 &  ---  &  66.1$\pm$11.9 &      ---      &      ---      &      ---      &       ---      &  2$\times$950 &  25 & 2 \\
         & 1155 & 06:50:09.5 & +05:44:28 & 15.754 & 1.729 &  ---  &  71.0$\pm$7.9  & 1.17$\pm$0.13 & 3.39$\pm$0.12 & 2.74$\pm$0.12 &  7.30$\pm$0.21 &  2$\times$950 &  29 & 1 \\
         & 1231 & 06:50:08.6 & +05:44:07 & 16.702 & 1.819 &  ---  &  66.1$\pm$7.7  & 1.13$\pm$0.14 & 3.21$\pm$0.21 & 2.77$\pm$0.10 &  7.11$\pm$0.27 & 2$\times$1500 &  25 & 1 \\
         & 1288 & 06:50:05.7 & +05:43:53 & 15.500 & 1.735 &  ---  &  68.3$\pm$7.2  & 1.27$\pm$0.11 & 3.51$\pm$0.12 & 2.93$\pm$0.12 &  7.71$\pm$0.20 &  2$\times$850 &  29 & 1 \\
         & 1421 & 06:50:18.4 & +05:43:22 & 15.556 & 2.063 &  ---  &  68.3$\pm$6.7  & 1.45$\pm$0.08 & 3.94$\pm$0.09 & 2.62$\pm$0.07 &  8.01$\pm$0.14 &  2$\times$900 &  40 & 1 \\
         & 1650 & 06:50:14.9 & +05:42:17 & 14.974 & 2.092 &  ---  &  97.1$\pm$7.8  &      ---      &      ---      &      ---      &       ---      &  2$\times$850 &  45 & 9 \\
Be~70    & 1105 & 05:25:44.8 & +41:56:44 & 12.651 & 2.035 & 2.014 & -13.8$\pm$7.6  & 1.94$\pm$0.05 & 4.66$\pm$0.05 & 3.50$\pm$0.06 & 10.10$\pm$0.09 &  2$\times$700 & 186 & 1 \\
         & 1088 & 05:25:45.0 & +41:55:55 & 14.097 & 1.994 & 1.849 &  -4.3$\pm$7.5  & 1.61$\pm$0.02 & 4.32$\pm$0.02 & 3.20$\pm$0.03 &  9.13$\pm$0.04 &  2$\times$800 &  98 & 1 \\
         & 0820 & 05:25:49.9 & +41:56:51 & 14.482 & 1.944 & 1.816 & -17.3$\pm$7.9  & 1.53$\pm$0.04 & 4.18$\pm$0.03 & 2.99$\pm$0.04 &  8.70$\pm$0.06 &  2$\times$800 &  69 & 1 \\
         & 0609 & 05:25:54.6 & +41:58:22 & 14.608 & 1.734 & 1.585 & -11.7$\pm$7.2  & 1.56$\pm$0.04 & 3.69$\pm$0.05 & 2.80$\pm$0.06 &  8.05$\pm$0.09 &  2$\times$850 &  44 & 1 \\
         & 0811 & 05:25:50.1 & +41:57:46 & 14.774 & 1.796 & 1.645 & -19.7$\pm$7.7  & 1.46$\pm$0.08 & 3.87$\pm$0.07 & 2.94$\pm$0.08 &  8.27$\pm$0.12 &  2$\times$850 &  58 & 1 \\
         & 0690 & 05:25:52.7 & +41:57:13 & 14.923 & 1.617 & 1.457 & -20.4$\pm$7.3  & 1.49$\pm$0.05 & 3.71$\pm$0.07 & 2.71$\pm$0.06 &  7.91$\pm$0.11 &  2$\times$950 &  58 & 1 \\
         & 0735 & 05:25:51.9 & +41:57:35 & 15.142 & 1.762 & 1.624 & -19.7$\pm$7.9  & 1.46$\pm$0.08 & 3.89$\pm$0.08 & 2.80$\pm$0.10 &  8.15$\pm$0.15 &  2$\times$950 &  40 & 1 \\
         & 0171 & 05:26:03.3 & +41:58:54 & 15.304 & 1.716 & 1.547 & -12.9$\pm$8.9  & 1.35$\pm$0.12 & 4.01$\pm$0.13 & 2.35$\pm$0.13 &  7.71$\pm$0.22 &  2$\times$900 &  30 & 1 \\
NGC~1798 & 0005 & 05:11:36.7 & +47 41 49 & 14.380 & 1.633 & 1.483 &   4.3$\pm$7.3  & 1.54$\pm$0.06 & 3.79$\pm$0.07 & 2.91$\pm$0.07 &  8.24$\pm$0.12 &  2$\times$800 &  56 & 1 \\
         & 0009 & 05:11:42.8 & +47:41:32 & 14.777 & 1.653 & 1.486 &   0.2$\pm$7.3  & 1.31$\pm$0.10 & 3.45$\pm$0.10 & 2.60$\pm$0.08 &  7.36$\pm$0.16 &  2$\times$750 &  40 & 1 \\
         & 0011 & 05:11:37.0 & +47:40:15 & 15.305 & 1.666 & 1.529 &  -3.4$\pm$7.0  & 1.40$\pm$0.09 & 3.85$\pm$0.11 & 2.71$\pm$0.08 &  7.96$\pm$0.16 &  2$\times$800 &  41 & 1 \\
         & 0013 & 05:11:41.2 & +47:40:41 & 15.325 & 1.660 & 1.524 &  -5.6$\pm$7.2  & 1.38$\pm$0.07 & 3.52$\pm$0.10 & 2.64$\pm$0.13 &  7.54$\pm$0.18 &  2$\times$850 &  40 & 1 \\
         & 0043 & 05:11:47.9 & +47:40:26 & 12.875 & 2.252 & 1.970 &  11.2$\pm$7.8  & 1.89$\pm$0.02 & 4.68$\pm$0.03 & 3.44$\pm$0.03 & 10.01$\pm$0.05 &  2$\times$700 & 110 & 1 \\
         & 0608 & 05:11:41.0 & +47:40:43 & 16.653 & 0.909 & 0.816 &  12.8$\pm$11.0 &      ---      &      ---      &      ---      &       ---      &  2$\times$850 &  25 & 2 \\
NGC~2266 & 0034 & 06:43:16.0 & +26 57 45 & 12.454 &  ---  & 0.957 &  21.0$\pm$7.7  &      ---      &      ---      &      ---      &       ---      &  2$\times$350 &  55 & 9 \\
         & 0067 & 06:43:20.2 & +26:57:32 & 10.538 &  ---  & 1.288 &  30.2$\pm$7.2  &      ---      &      ---      &      ---      &       ---      &   2$\times$50 &  55 & 9 \\
         & 0075 & 06:43:12.9 & +26:57:07 & 15.242 &  ---  & 0.341 & -18.0$\pm$9.2  &      ---      &      ---      &      ---      &       ---      &  2$\times$350 &  24 & 2\\
         & 0096 & 06:43:28.0 & +26:58:10 & 12.287 &  ---  & 1.253 &  -8.9$\pm$7.5  & 1.47$\pm$0.05 & 3.84$\pm$0.05 & 2.65$\pm$0.06 &  7.96$\pm$0.09 &  2$\times$250 &  58 & 1 \\
         & 0101 & 06:43:25.3 & +26:57:49 & 12.978 &  ---  & 1.187 & -19.1$\pm$7.8  & 1.46$\pm$0.06 & 3.71$\pm$0.06 & 2.25$\pm$0.10 &  7.42$\pm$0.13 &  2$\times$500 &  50 & 1 \\
         & 1011 & 06:43:25.3 & +26:57:50 & 12.117 &  ---  & 1.273 & -18.5$\pm$7.1  & 1.43$\pm$0.06 & 4.13$\pm$0.05 & 2.97$\pm$0.06 &  8.53$\pm$0.10 &  2$\times$500 &  58 & 1\\       \hline\hline
\end{tabular}
\tablefoot{ (1)  member star used both for radial velocity and metallicity determination; (2) member star used only for radial velocity determination; (9) non-member star. }
\end{scriptsize}
\end{table*}

The goal of this paper is to increase the number of clusters with radial
velocities and metallicities determined from spectroscopy. For this purpose, we
have selected four OCs, Berkeley~ 26, Berkeley~70, NGC~1798, and NGC~2266, that
had not
been studied spectroscopically before. These clusters are located towards the
Galactic anticenter, at relatively  large distances (R$_{GC}\geq$10 kpc) where
the trend of the radial metallicity seems to flatten
\citepads[e.g.][]{2011A&A...535A..30C}. Medium-resolution spectra in the region
of the near-infrared \ion{Ca}{II} triplet (CaT) at $\sim$8500\AA~were obtained
for several red giant branch (RGB) stars in each cluster. The
observations and data reduction are described in Sect.~\ref{sec2}, radial
velocities and metallicities are obtained in Sects.~\ref{sec3}--\ref{sec4}, the
results for each cluster are discussed and compared with the literature in
Sect.~\ref{sec5}, and the comparison of the results obtained here with the
trends observed in the disk is performed in Sec.~\ref{sec6}. Finally, our main
conclusions are summarised in
Sect.~\ref{sec7}.

\section{Observations and data reduction}\label{sec2}

\begin{table}
\begin{minipage}[htb]{\columnwidth}
\caption{Adopted cluster parameters.}
\label{clusterspropierties}
\centering
\renewcommand{\footnoterule}{}  
\begin{tabular}{l c c c c c}
\hline\hline       
Cluster   & E(B--V) & (m-M)$_o$ & Age & R$_{gc}$ & z \\
          & (mag)   & (mag)     & (Gyr)  & (kpc) & (kpc) \\
\hline
Be~26\footnote{Average of the values by \citetads{2010MNRAS.402.2720P} and  \citetads{2008PASJ...60.1267H}. See text for details.}
          & 0.68$\pm$0.06 & 13.8$\pm$0.5 & 4.25$\pm$0.25 & 13.9 & 0.24\\
Be~70\footnote{Values taken from \citetads{2002AJ....123..905A} which are similar to those derived by \citetads{2004PASJ...56..295H} with the exception that they derived an age of 2.8 Gyr. }
          & 0.48$\pm$0.10 & 13.1$\pm$0.3 & 4.0$\pm$0.8 & 12.6 & 0.26 \\	 
NGC~1798\footnote{Values taken from \citetads{1999MNRAS.304..883P} which are similar to others available in the literature within error bars \citepads[e.g.][]{2000A&A...359..347D,2002NewA....7..553T,2004A&A...414..163S,2007A&A...467.1065M}.}
          & 0.51$\pm$0.04 & 13.1$\pm$0.2 & 1.4$\pm$0.3 & 12.5 & 0.35 \\
NGC~2266\footnote{Values taken from \citetads{2012A&A...539A.125D}. Similar values have been obtained by other investigations in this cluster \citepads{1991AcA....41..191K,2000A&A...359..347D,2001NewA....6..293T,2002NewA....7..553T,2002A&A...388..158L,2004A&A...414..163S,2006MNRAS.371.1641P,2007A&A...467.1065M,2008BaltA..17...51M,2010A&A...517A..32P,2010A&A...516A...2M}}
          & 0.15$\pm$0.01  & 12.6$\pm$0.2 & 0.7$\pm$0.1 & 11.8 & 0.59 \\
\hline
\end{tabular}
\end{minipage}
\end{table}

\begin{figure}
\centering
\includegraphics[height=14cm,width=\columnwidth]{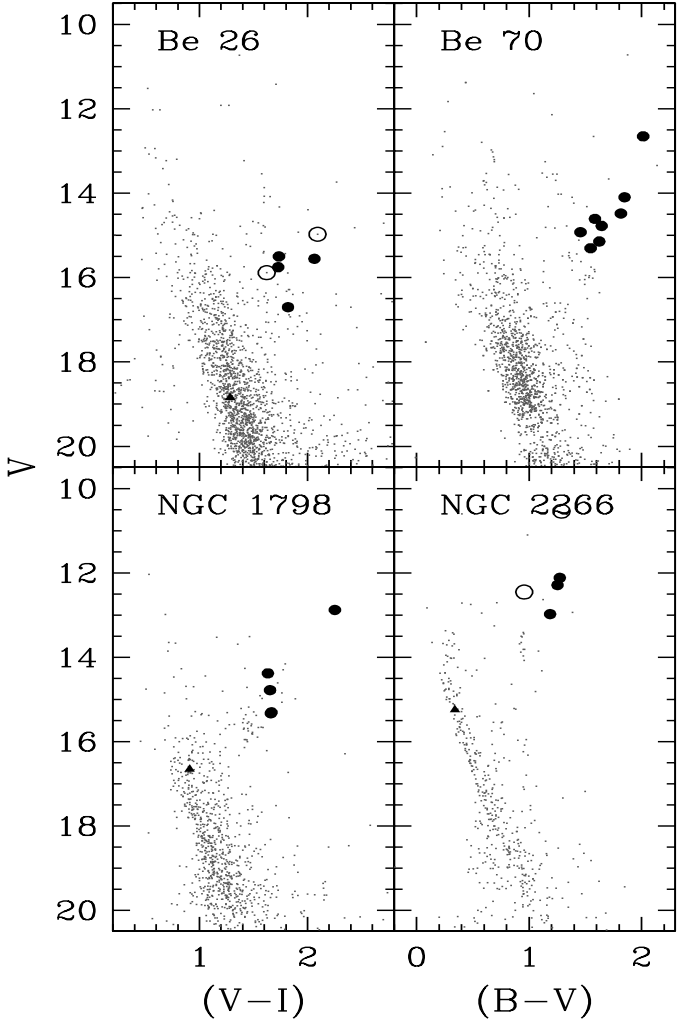}
\caption{Color-magnitude diagrams of clusters studied in this paper (small gray
dots). Filled circles are stars assumed to be cluster members based on their
radial velocities and used in the metallicity determination. Filled triangles
are main-sequence stars assumed to be members based on their radial velocity but not
used to determine the metallicity of these clusters. Open circles are
stars rejected as possible members based on their radial velocities.}
\label{CMDs}%
\end{figure}

The target stars were selected from the color-magnitude diagram of each cluster
taken from \citetads{2010MNRAS.402.2720P}, \citetads{2002AJ....123..905A},
\citetads{1999MNRAS.304..883P}, and \citetads{2007A&A...467.1065M} for
Berkeley~26, Berkeley~70, NGC~1798, and NGC~2266, respectively
(Fig.~\ref{CMDs}). These data together with the coordinates of each target stars
were obtained from the WEBDA\footnote{{\tt http://www.univie.ac.at/webda}}
database \citepads{1995ASSL..203..127M}. In total, we observed 26 stars: 8 in
Berkeley~70, and 6 in the others. Table~\ref{obsstars} summarizes the coordinates, 
magnitudes, exposure times, and signal--to--noise  ratios for each target star.
The global properties of each cluster are listed in
Table~\ref{clusterspropierties}.

Observations were carried out between the nights of 5 and 9 of February 2012
using the Intermediate Dispersion Spectrograph (IDS) mounted at the Cassegrain
focus of the 2.5 m Isaac Newton Telescope (INT) located at the Roque de los
Muchachos Observatory, Spain. We used the R1200R grism centered at 8500\AA~and
the RED+2 CCD, providing a spectral resolution of about 8000. For each target, we
obtained two exposures with the star shifted along the slit in each of them. In
some cases, another star was observed together with the main target because it
was aligned with the slit. Since stellar spectra have reasonable signal-to-noise ratios, we
also analyzed these stars. In all cases, these are main-sequence stars, which
therefore, were used only for the radial velocity analysis.

The data reduction was performed using \textit{imred} and \textit{specred}
packages within IRAF\footnote{Image Reduction and Analysis Facility, IRAF is
distributed by the National Optical Astronomy Observatories, which are operated
by the Association of Universities for Research in Astronomy, Inc., under
cooperative agreement with the National Science Foundation.}. Firstly, each
image was overscan-subtracted and flat-field corrected. Then, since we had acquired
two images of each target with the star shifted along the slit, we subtracted one
from the other, obtaining a positive and a negative spectrum in the same image.
With this procedure, the sky was subtracted from the same physical pixel in which
the star was observed, thus minimizing the effects of pixel-to-pixel sensitivity
variations. A time dependency of course remained, since the two spectra had not
been taken simultaneously. These sky residuals were eliminated in the following
step, in which the spectrum was extracted in the traditional way and the
remaining sky background subtracted from the information on both sides of the
stellar spectra. In the next step, the spectrum was wavelength-calibrated. We then
again subtracted the negative from the positive spectrum (so we effectively added both spectra
because one is negative) to obtain the final spectrum. Finally, each spectrum
was normalized by fitting a polynomial, excluding the strongest lines in the
wavelength range, such as those of the CaT.

\section{Radial velocities}\label{sec3}

\begin{figure}
\centering
\includegraphics[height=\columnwidth,width=\columnwidth]{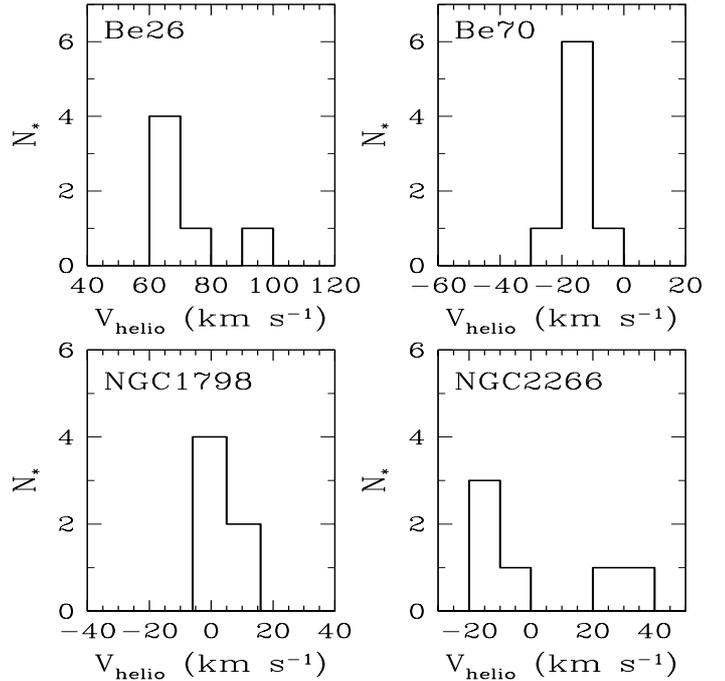}
\caption{Radial velocity distributions of the four clusters studied.}
\label{Vrdist}%
\end{figure}

The radial velocity of each star was calculated using the \textit{fxcor}
task in IRAF, which performs a cross-correlation between the target and template
spectra of known radial velocity. As templates, we used two bright stars, 
HD15656 and HD74462, observed in the same nights and with a high
signal--to--noise ratio. The final radial velocity for each target star was
obtained as the average of the velocities obtained for each template, weighted
by the width of the correlation peaks. The measured  radial velocities for each star are
listed in column 8 of Table~\ref{obsstars}. The velocity distribution of each
cluster is shown in Fig.~\ref{Vrdist} and discussed in Sec.~\ref{sec5}.

\section{CaT equivalent widths and metallicity determination}\label{sec4}

The equivalent widths of the CaT lines of the observed RGB stars were used
to determine the metallicity of each cluster. This technique was originally
developed to determine the metallicities of old and metal-poor populations such as
those of globular clusters
\citepads[e.g.][]{1988AJ.....96...92A,1991AJ....101.1329A,1997PASP..109..907R}.
However, more recent investigations have extended its validity to much larger
ranges of ages and metallicities, including extremely metal-poor regimes
\citepads{2010A&A...513A..34S} and the young and metal-rich OC populations
\citepads{2004MNRAS.347..367C,2007AJ....134.1298C}.

Although the procedure used to determine metallicities from the CaT lines
strengths is described in depth by \citetads{2007AJ....134.1298C}, we summarize
here the main steps. Firstly, we determine the equivalent width of each line by
fitting its profile with a Gaussian plus a Lorentzian. This combination provides
the best fit to the line core and wings. Although in our case the spectra had already been
normalized, the position of the continuum was recalculated by performing a linear fitting
to the mean values of each continuum bandpass. The bandpasses used to fit the
line profile and determine the continuum position are those described by
\citetads{2001MNRAS.326..959C}. The CaT index, denoted $\Sigma$Ca, is defined as the sum of the equivalent widths of the three CaT
lines. We list the equivalent widths of the three CaT lines and the $\Sigma$Ca values obtained for each target star in
Table~\ref{obsstars}.

The strengths of the CaT lines do not only depend on the chemical abundance,
but also on the  temperature and gravity of the stellar atmosphere. Since we are
only interested in the abundance dependence, the temperature and gravity
dependences must be removed. To do this, we used the finding that, for a given
chemical abundance, the stars define a sequence in the luminosity-$\Sigma$Ca
plane when the temperature and/or gravity change. It is known that this sequence
is not linear for extremely metal-poor regimes
\citepads{2010A&A...513A..34S} and in the bottom of the RGB
\citepads{2007AJ....134.1298C}. However, we can neglect this trend since our
targets are located in the upper part of the RGB and because for OCs we expect
metallicities of around solar. Traditionally, the difference between the
$V$ magnitude of a star and that of the Horizontal Branch (HB), $V-V_{HB}$,
has been used as a luminosity indicator, which also removes any dependence on
distance and reddening
\citepads[e.g.][]{1991AJ....101.1329A,1997PASP..109..907R,2004MNRAS.347..367C}.
However, in many stellar populations, it is difficult to define accurately the
position of the HB, such as poorly populated OCs. For this reason, other authors
have used the absolute magnitude in either the $V$ or $I$ bandpasses
\citepads[e.g.][]{1991AJ....101.1329A,2004AJ....127..840P,2007AJ....134.1298C}.
We used this second approach in this paper. The CaT metallicity calibration
based on M$_I$ is less sensitive to age. Unfortunately, $I$ magnitudes are unavailable for the stars of one of the clusters, NGC~2266. Therefore, we derive the metallicities of each cluster using both M$_V$ and M$_I$ to minimize the effect of age on the metallicity determination of NGC~2266.
 
\begin{figure}
\centering
\includegraphics[height=7cm,width=\columnwidth]{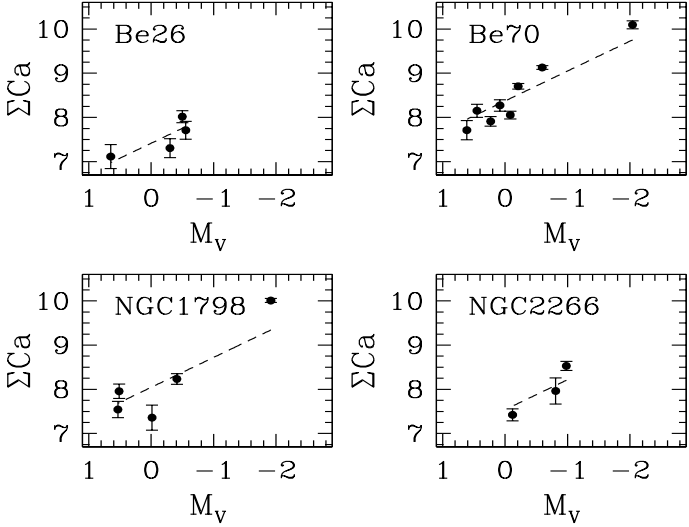}
\caption{Observed stars in each cluster in the M$_V$-$\Sigma$Ca plane. Dashed lines are the linear fit to the stars in each cluster.}
\label{MvSigmaCa}%
\end{figure}

\begin{figure}
\centering
\includegraphics[height=7cm,width=\columnwidth]{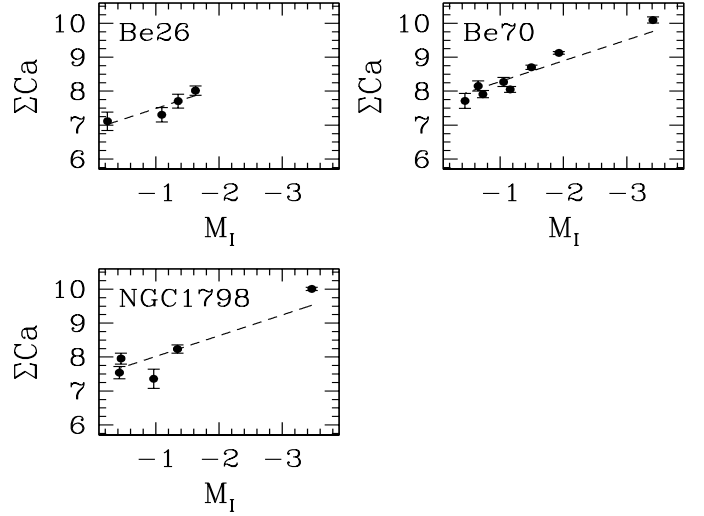}
\caption{Same as Fig.~\ref{MvSigmaCa} but for M$_I$-$\Sigma$Ca plane. }
\label{MiSigmaCa}%
\end{figure}

In the M$_{V,I}$-$\Sigma$Ca planes, the sequences can be parametrized in the form
$\Sigma Ca = W'_{V,I} + \beta_{V,I}\times M_{V,I}$. The slopes $\beta_{V,I}$ are
independent of metallicity as demonstrated by
\citetads{2007AJ....134.1298C}, whereas $W'_{V,I}$, the so-called reduced
equivalent width, varies as a function of the chemical abundance. We used the
slopes $\beta_V=-0.677\pm0.004$ \AA~mag$^{-1}$ and $\beta_I=-0.611\pm0.002$ 
\AA~mag$^{-1}$ calculated by \citetads{2007AJ....134.1298C}. The position of the
observed stars in each cluster in the M$_V$-$\Sigma$Ca and M$_I$-$\Sigma$Ca
planes are shown in Fig.~\ref{MvSigmaCa} and \ref{MiSigmaCa}, respectively. The
$W'_{V,I}$ values obtained for each cluster are listed in
Table~\ref{clustersresults}.

Relationships between $W'_{V,I}$ and metallicity on different metallicity
scales, including the widely used one from \citetads{1997A&AS..121...95C} were
derived by \citetads{2007AJ....134.1298C}. \citetads{2009A&A...508..695C} updated their metallicity scale based on more
homogeneous and higher resolution spectra of a larger sample of bright giants in
24 Galactic globular clusters. They provided a detailed comparison between their
old and new scales. The existence of this new metallicity scale for globular
clusters and the new determinations of metallicities available for some open clusters used
by \citetads{2007AJ....134.1298C} to derive their calibrations, motivated
\citetads{2011AJ....142...61C} to obtain new relationships between $W'_{V,I}$
and metallicity. 
We used these updated relations to obtain the metallicity, on the
\citetads{2009A&A...508..695C} scale, of the clusters studied in this paper. The
values obtained for each cluster are listed in Table~\ref{clustersresults} and
discussed below. We have to note that the metallicities derived are
related to the reddenings and distance moduli used as references in each case.
For these reason, the metallicity error bars were estimated by taken into
account the uncertainties in these quantities.

\section{Cluster-by-cluster discussion}\label{sec5}

\begin{table*}
\begin{minipage}[t]{17.5cm}
\caption{Cluster metallicities and radial velocities}
\label{clustersresults}
\centering
\renewcommand{\footnoterule}{}  
\begin{tabular}{l c c c c c}
\hline\hline       
Cluster   & W'$_V$ & [Fe/H]$_V$ & W'$_I$ & [Fe/H]$_I$ & $\langle V_r \rangle$ \\
          & (mag)   & (dex)     & (mag)  & (dex) & (km s$^{-1}$) \\
\hline
Be~26 & 7.42$\pm$0.20 & -0.43$\pm$0.20 & 6.88$\pm$0.23 & -0.35$\pm$0.17 & 68$\pm$12\\
Be~70 & 8.37$\pm$0.01 & -0.02$\pm$0.18 & 7.67$\pm$0.01 & -0.01$\pm$0.14 & -15$\pm$7\\	 
NGC~1798 & 8.05$\pm$0.04 & -0.16$\pm$0.10 & 7.41$\pm$0.03 & -0.12$\pm$0.07 & 2$\pm$10\\
NGC~2266 & 7.54$\pm$0.02  & -0.38$\pm$0.07 & --- & --- & -16$\pm$15\\
\hline
\end{tabular}
\end{minipage}
\end{table*}

\subsection{Berkeley~26}

Also known as Biurakan~12, Berkeley~26 is a poorly studied cluster in Monoceros.
To our knowledge, there are only three studies of this cluster in the
literature. \citetads{2008PASJ...60.1267H} and \citetads{2010MNRAS.402.2720P}
obtained the color-magnitude diagram of Berkeley~26 in the BVI photometric
system. \citetads{2008MNRAS.389..285T} analyzed near-infrared JHK photometry
obtained from 2MASS\footnote{2MASS (Two Mi\-cron All Sky Survey) is a
joint project of the University of Massachusetts and the Infrared Processing and
Analysis Center/California Institute of Technology, funded by the National
Aeronautics and Space Administration and the National  Science Foundation.}. All
of these authors derived the age, reddening, distance, and metallicity by means of
isochrone fitting. There are some discrepancies between the values obtained by each of
them. There is good agreement, within the uncertainties, in the values derived
by \citetads{2008PASJ...60.1267H} and \citetads{2010MNRAS.402.2720P}, but not
between them and the \citetads{2008MNRAS.389..285T} measurements. For example,
\citetads{2008PASJ...60.1267H} and \citetads{2010MNRAS.402.2720P} derived an age
of 4.5 Gyr and  4 Gyr, respectively. In contrast, \citetads{2008MNRAS.389..285T}
obtained a much younger age of 0.6 Gyr. A reddening of E(B-V)=0.61 was derived
by \citetads{2008PASJ...60.1267H}, which is between the values derived by
\citetads{2008MNRAS.389..285T} of E(B-V)=0.5, and \citetads{2010MNRAS.402.2720P}
of E(B-V)=0.75. In the case of the distance modulus, the obtained values range
from (m-M)$_0$=12.6 \citepads{2008MNRAS.389..285T} to 14.45
\citepads{2008PASJ...60.1267H}. \citetads{2010MNRAS.402.2720P} derived a value
between both of (m-M)$_0$=13.17. As was noted by
\citetads{2010MNRAS.402.2720P},  the poor quality of the data used by \citetads{2008MNRAS.389..285T}, mainly owing to the techniques that they used to decontaminate the foreground
contribution and select the cluster members, may explain the discrepancies with other investigations in the literature. For this reason, we used in our
work the distance modulus and reddening obtained by averaging the values derived
by \citetads{2008PASJ...60.1267H} and \citetads{2010MNRAS.402.2720P} listed in
Table~\ref{clusterspropierties}.

A total of six stars were observed along the Berkeley~26 line of sight. The
spectrum of the star 1650 contains clear molecular bands. Moreover, this star has a larger
radial velocity than the other observed stars. For these reasons, we use
this star in neither the radial velocity nor the metallicity determinations.
The star 1038 is not a RGB and was used only for the radial velocity
determination. Therefore, excluding star 1650, we derived a mean radial velocity
for this cluster of $\langle V_r \rangle$=68$\pm$12  km s$^{-1}$

Only \citetads{2010MNRAS.402.2720P} provided an estimation of the Berkeley~26
metallicity of [Fe/H]=-0.70 dex. \citetads{2008PASJ...60.1267H} assumed a
metallicity of [Fe/H]$\sim$-0.35 for this cluster. We derived
[Fe/H]=-0.43$\pm$0.20 and  -0.35$\pm$0.17 dex using M$_V$ and M$_I$,
respectively. These values are in relatively good agreement with the photometric
estimations taken into account the uncertainties. As explained above, 
\citetads{2008MNRAS.389..285T} derived a different distance modulus and reddening
for Berkeley~26 than those used here. When these values were used in the
metallicity determination, we obtained [Fe/H]=+0.09 and +0.06 in the $V$ and $I$
bandpasses, respectively. 

\subsection{Berkeley~70}

Berkeley~70 is a rich OC, which probably belongs to the Perseus arm
\citepads{2008A&A...482..777C}. It shows a highly populated main-sequence and a
well-developed RGB. A significant amount of blue stranglers can also be
observed. To our knowledge, only two studies have determined the properties of
this cluster, using both $UBV$ photometry
\citepads{2002AJ....123..905A,2004PASJ...56..295H}. The reddenings and distances
determined by them are the same within the uncertainties. There is a slight
difference in the ages obtained, although they are consistent with the uncertainties.
While \citetads{2002AJ....123..905A} determined an age of 4 Gyr,
\citetads{2004PASJ...56..295H} obtained 2.8 Gyr. 

The eight observed stars have  similar radial velocities so we conclude that all
of them are members of Berkeley~70. They define a narrow velocity
distribution (top-right panel of Fig.~\ref{Vrdist}). We derive a mean radial
velocity for this clusters of $\langle V_r \rangle$=-15$\pm$7  km s$^{-1}$.

For Berkeley~70, we derived a metallicity of [Fe/H]=-0.02$\pm$0.18 and
-0.01$\pm$0.14 dex using $V$ and $I$ magnitudes, respectively. To our knowledge,
the only determinations of metallicity for this cluster available in the
literature come from isochrone fitting, and they are all lower than obtained here. \citetads{2002AJ....123..905A} determined a metallicity
of -0.32 dex, while \citetads{2004PASJ...56..295H} obtained a slightly more
metal-poor value of -0.48 dex. This discrepancy may be explained by the
large uncertainties involved in the metallicity estimation from isochrone fitting owing to the limitation of the method itself and to the sparse color magnitude-diagram used as a reference.

\subsection{NGC~1798}

NGC~1798, also denoted as Berkeley~16, is another poorly studied OC located in
the Auriga constellation. Its color-magnitude diagram shows a well-populated
main-sequence and red clump, and a poorly defined RGB. The reddening towards
the NGC~1798 is relatively high, E(B-V)=0.51$\pm$0.04
\citepads{1999MNRAS.304..883P,2000A&A...359..347D,2002NewA....7..553T,
2004A&A...414..163S,2007A&A...467.1065M}, which is in part explained by its
proximity to the Galactic plane. An age of 1.4$\pm$0.2 Gyr has been estimated by
most of the previous investigations in this cluster
\citepads[e.g.][]{1994AJ....108.1773J,1999MNRAS.304..883P,2002A&A...388..158L,
2004A&A...414..163S,2007A&A...467.1065M}. These works have also obtained a similar
distance modulus for this cluster of (m-M)$_0$=13.1$\pm$0.2.

The six stars observed in the NGC~1798 field seem to be members of this cluster
based on their radial velocity. The main-sequence star 608 was observed because it
was aligned with the slit in the observations of star 13 but it was only used for the radial velocity analysis. We determined a mean radial velocity of
$\langle V_r \rangle$=2$\pm$10 km s$^{-1}$ for this cluster.

For the five RGB stars observed in NGC~1798, we inferred metallicities of [Fe/H]=-0.16$\pm$0.10 and
-0.12$\pm$0.07 dex using $V$ and $I$ magnitudes, respectively. There are a
handful of metallicity estimations for NGC~1798 in the literature. All of them
assign a metallicity lower than that obtained here, [Fe/H]=-0.46$\pm$0.10 dex
\citepads{1999MNRAS.304..883P,2001NewA....6..293T,2003NewA....8..737T,
2010A&A...517A..32P}.  As in the case of Berkeley~70, this disagreement may be explained by the
large uncertainties in the metallicities derived from isochrone fitting.

\subsection{NGC~2266}

NGC~2266, also known as Melotte~50, is a rich and well-condensed system. It is
the youngest cluster of our sample with an age similar to that of the Hyades,
\citepads[0.7$\pm$0.1 Gyr;
][]{1991AcA....41..191K,2004A&A...414..163S,2010A&A...516A...2M,
2012A&A...539A.125D}, although some authors have obtained a slightly older age
\citepads[1.2$\pm$0.1 Gyr; ][]{2007A&A...467.1065M,2008BaltA..17...51M}.  Its
color-magnitude diagram has a populated main-sequence, especially in the turn-off
area, and a well-defined red-clump located at $V\sim$13.7 with a few stars above
it (Fig.~\ref{CMDs}). The contamination of the color-magnitude diagram seems
to be almost negligible. We used the reddening and distance derived
by \citetads{2012A&A...539A.125D} (see Table~\ref{clusterspropierties}),
which are similar, within the uncertainties to others available in the
literature
\citepads[e.g.][]{2002A&A...388..158L,2006MNRAS.371.1641P,2008BaltA..17...51M,
2010A&A...516A...2M}.

Six stars were observed in this cluster. However, it seems that only four of
them, including the main-sequence object 75, are true members of the cluster.
The other two, stars 34 and 67, have a radial velocity differing from the average of the other stars much more than the expected uncertainties, hence may not be members of this cluster. Excluding
these two stars, we obtained a mean radial velocity of $\langle V_r \rangle$=-16$\pm$15  km
s$^{-1}$ for NGC~2266.

Since one of the four stars confirmed as a NGC~2266 member is a main-sequence
object, its metallicity determination is based on only three stars. Moreover,
$I$ magnitudes are unavailable for these stars. We therefore, obtained a
metallicity of [Fe/H]=-0.38$\pm$0.07 dex derived only from $V$ magnitudes. This
value is in very good agreement with the following estimates available in the literature:
-0.26$\pm$0.20 dex \citepads{2010A&A...517A..32P}; -0.68
\citepads{2008BaltA..17...51M}; -0.26$\pm$-0.18 \citepads{1991AcA....41..191K};
and -0.35 \citepads{2010A&A...516A...2M,2012A&A...539A.125D}. We note, however, that
\citetads{2004A&A...414..163S} assigned a more metal-rich metallicity of
0.00$\pm$0.20 dex.
\begin{figure}
\centering
\includegraphics[height=10cm,width=\columnwidth]{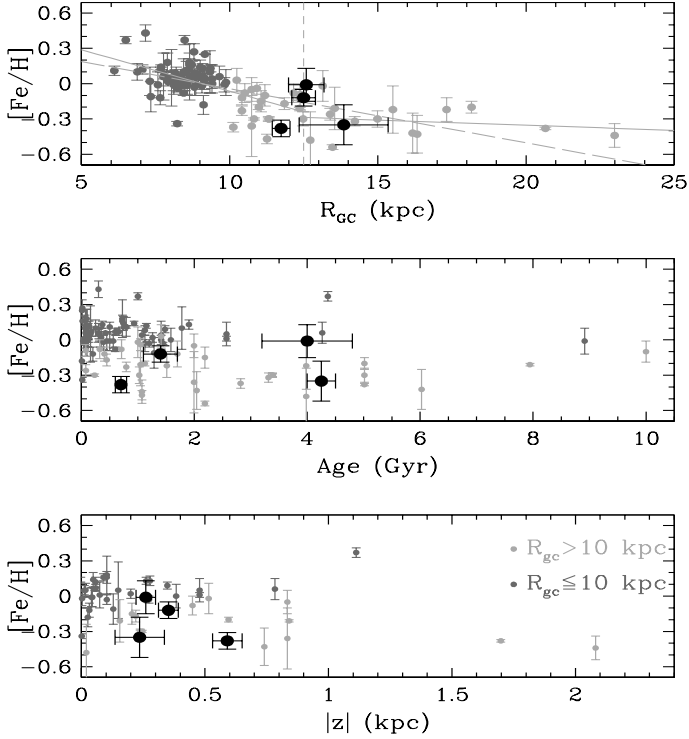}
\caption{Run of [Fe/H] with R$_{gc}$ (top), age (middle), and $|z|$ (bottom) of
the OCs in the \citetads{2011A&A...535A..30C} compilation. Dark and light gray
points are clusters in the inner 10 kpc and from there off, respectively. Black
points are the clusters studied in this paper.}
\label{trends}%
\end{figure}

\section{Comparison with trends in the Galactic disk}\label{sec6}

As mentioned earlier, OCs are fundamental test particles in the study of the
formation and evolution of the Galactic disk. In particular, they are used to
trace the existence of trends in metallicity with Galactocentric distance,
Rgc, the vertical distance to the Galactic plane, z, and age, providing
fundamental constraints on chemical evolution models. All the investigations
based on OCs agree that the iron content decreases almost linearly
with increasing radius
\citepads[e.g.][]{2002AJ....124.2693F,2010A&A...511A..56P}. However,
investigations based on samples containing clusters at larger distances
 found that the metallicity decreases as a function of
increasing radius to R$_{gc}\sim$12.5 kpc and appears to flatten from there
outwards \citepads[e.g.][]{2005AJ....130..597Y,2008A&A...488..943S,2011MNRAS.412.1265A,
2011A&A...535A..30C}. Moreover, it seems that this gradient has evolved with time being, having been 
steeper in the past
\citepads[e.g.][]{2009A&A...494...95M,2011MNRAS.412.1265A,2011A&A...535A..30C}.
However, there are no hints of the existence of an age-metallicity relationship
\citepads[e.g.][]{1995ARA&A..33..381F}. This indicates that the chemical
enrichment of OCs is modulated by their location in the Galaxy and not by the
moment at which they formed \citepads{2011A&A...535A..30C}. 

It is helpful to check how the clusters studied in this paper fit the general trends
observed in the Galactic disk. This comparison, of course, should be interpreted with caution
because the results obtained here should be confirmed by analyses based on
high-resolution spectra. In all cases, this comparison is
interesting because of the location of these clusters at relatively large
distances from the Galactic anticenter (R$_{gc}<$10 kpc). To perform this comparison, we used the OC compilation
obtained by \citetads{2011A&A...535A..30C}, which contains data for 89 clusters whose
chemical abundances were derived by investigations based on high-resolution spectra
(R$\ge$20000) available in the literature. We refer the reader to this paper for
details. The run of metallicity with R$_{gc}$, age, and $|z|$ for this sample
have been plotted in the top, middle, and bottom panels of Fig.~\ref{trends},
respectively. The solid lines in the top panel are the linear fits to the sample
assuming a change in the slope at R$_{gc}\sim$12.5 kpc, while the long-dashed
line is the linear fit to the whole sample. Clusters in the inner 10 kpc and
from there outwards have been plotted with dark and light gray circles, respectively, in
the three panels.

The four clusters studied here have been overplotted with
black symbols. Berkeley~26, in spite of the uncertainty in its distance, and NGC~1798 closely follow the general trends
observed in other coeval clusters located at similar Galactocentric distances.
NGC~2266 seems to be slightly younger than other metal-poor systems of similar metallicity situated at the
same distance. Nevertheless, its metal content is akin to other OCs located away from the Galactic plane, such as NGC~2266. In all cases, this comparison must be considered be taken with caution because its
metallicity has been obtained from only three stars. In contrast, Berkeley~70,
which has a more robust metallicity determination based on six stars, appears more metal-rich than
other systems located at $\sim$12.5 kpc. There are a few clusters with similar ages
and metallicities to Berkeley~70, but all of them are situated in the inner 10
kpc at the same distance. Although it is beyond the scope of this paper, one explanation could be that this cluster was formed at a smaller
Galactocentric distance and then migrated to its current location. However, more information than is currently available is needed to confirm or disprove this hypothesis.

\section{Conclusions}\label{sec7}

We have analyzed medium-resolution spectra (R$\sim$8000) in the infrared CaT region ($\sim$8500 \AA) of several stars in four OCs: Berkeley~26, Berkeley~70, NGC~1798, and NGC~2266. To our knowledge, this is the first time that these clusters have been studied spectroscopically. Our
main results can be summarised as follows:

\begin{itemize}
 \item For Berkeley~26, we derived a mean radial velocity of $\langle V_r \rangle$=68$\pm$12 km s$^{-1}$
based on six stars. One of them is a main-sequence star that was not used in the
metallicity determinations. Using $V$ and $I$ magnitudes, we determined a
metallicities of [Fe/H]=-0.43$\pm$0.20 and -0.35$\pm$0.17 dex, respectively.
\item The eight observed stars in Berkeley~70 were confirmed as members based on
their radial velocity. For them, we derived a mean radial velocity of $\langle V_r \rangle$=-15$\pm$7
km s$^{-1}$ and metallicities of [Fe/H]=-0.02$\pm$0.18 and -0.01$\pm$0.14 dex based on $V$ and
$I$ magnitude data, respectively.
\item We derived a mean radial velocity for NGC~1798 of $\langle V_r \rangle$=2$\pm$10 km s$^{-1}$
from six stars, although one object is on the main sequence. For the
RGB stars, we obtained a metallicity of [Fe/H]=-0.16$\pm$0.10 and -0.12$\pm$0.07
dex in $V$ and $I$ bandpasses, respectively.
\item In the case of NGC~2266, its mean radial velocity, $\langle V_r \rangle$=-16$\pm$15 km s$^{-1}$
was obtained for four objects. A metallicity of [Fe/H]=-0.38$\pm$0.07 dex was
derived from the $V$ magnitudes of the three RGB stars observed in this cluster.
\end{itemize}

Finally, we investigated how the four analyzed clusters fit the trends
defined by other well-studied OCs. This comparison is motivated by our
clusters being situated at distances where other investigations have observed a
change in the slope of the metallicity gradient. In general, Berkeley~26 and
NGC~1798 follow the trends described by other coeval systems situated at the
same distance. In the case of NGC~2266, its height above the Galactic plane can
explain its low metallicity compared to other clusters of the same age.
In contrast, Berkeley~70 seems to be more metal-rich than other coeval clusters
situated at similar distances. We suggest that this cluster may have formed
at relatively small Galactocentric distances and has migrated outwards with time.
Nevertheless, more information is needed to confirm or discard this hypothesis. 

\begin{acknowledgements}
R. C acknowledge the support of the Spanish Ministry of Economy and Competitiveness
(Plan Nacional de Investigaci\'on Cient\'{\i}fica, Desarrollo, e Investigaci\'on
Tecnol\'ogica, AYA2010-16717). R. C. also acknowledges the
funds of the Spanish Ministry of Science and Innovation under the Juan de la
Cierva fellowship. This research has made use of the WEBDA database, operated at the Institute for Astronomy of the University of Vienna, and the SIMBAD database,
operated at CDS, Strasbourg, France
\end{acknowledgements}

\end{document}